# Domain evolution of BaTiO$_3$ ultrathin films under electric field: a first-principles study

Bo-Kuai Lai, Inna Ponomareva, Igor A. Kornev, L. Bellaiche and G. J. Salamo

Physics Department, University of Arkansas, Fayetteville, Arkansas 72701

**ABSTRACT**

A first-principles-derived method is used to study the morphology and electric-field-induced evolution of stripe nanodomains in (001) BaTiO$_3$ (BTO) ultrathin films, and to compare them with those in (001) Pb(Zr,Ti)O$_3$ (PZT) ultrathin films. The BaTiO$_3$ systems exhibit 180$^o$ periodic stripe domains at null electric field, as in PZT ultrathin films. However, the stripes alternate along [1-10] in BTO systems versus [010] in PZT systems, and no in-plane *surface* dipoles occur in BTO ultrathin films (unlike in PZT materials). Moreover, the evolution of the 180$^o$ stripe domains in the BaTiO$_3$ systems, when applying and increasing an electric field along [001], involves four regions: Region I for which the magnitude of the "down" dipoles (i.e., those that are antiparallel to the electric field) is reduced, while the domain walls do not move; Region II in which some local down dipoles adjacent to domain walls switch their direction, resulting in zigzagged domain walls − with the overall stripe periodicity being unchanged; Region III in which nanobubbles are created, then contract along [110] and finally collapse; and Region IV which is associated with a single monodomain. Such evolution differs from that of PZT ultrathin films for which neither Region I nor zigzagged domain walls exist, and for which the bubbles contract along [100]. Discussion about such differences is provided.

# I. INTRODUCTION

Ferroelectric materials are the heart of many applications such as non-volatile memories, communication devices, and microactuators. [1] The continuing miniaturization of these devices has stimulated considerable research attention on ferroelectric thin films (see, e.g., Refs. 2 and 3 and references therein). Such systems can exhibit striking phenomena because of some finite-size effect. For instance, they can adopt periodic stripe domains with exceptionally small periods (i.e., in the order of a few nanometers), when experiencing some specific mechanical and electrical boundary conditions. [2-12] Interestingly, the morphology of such nanostripe patterns seem to dramatically depend on the materials, as suggested by the facts that 180$^{\circ}$ nanostripes are known to alternate along the [010] direction in (001) Pb(Zr,Ti)O$_3$ (PZT) and PbTiO$_3$ thin films [5, 6, 9-11] while Ref. 4 predicts that such stripes alternate along another direction (namely, along [1-10]) in (001) BaTiO$_3$ (BTO) ultrathin films. The exact reason behind such difference, that is reminiscent of distinct behaviors found in Pb- *versus* Ba-based alloys, [13, 14] remains to be addressed. One may also wonder if other differences exist for the morphology of the stripe domains in PZT and BTO films, and, if so, why. For instance, do dipoles at the *surfaces* of BTO films also lie in-plane (to close the flux as in PZT films [6, 9] and in the Landau-Lifschitz model of domains [15]), or rather prefer to all lie along the growth direction (as in the Kittel model of domains [16]), or even adopt an intermediate case (e.g., form an angle of 45 degree between the in-plane and out-of-plane directions)? Another fundamental question to be addressed regarding stripe domains in BTO thin films is their precise atomistic evolution under the application of the factor that lies at the heart of many ferroelectric devices, namely an external electric field. More specifically, how do the differences between the initial stripe domains in BTO and PZT films affect the (very unusual) stripe → ferroelectric nanobubble → monodomain transition sequence that has been recently predicted and documented in details for PZT ultrathin films under applied fields [9]? In particular, it would be worthwhile to have a deep microscopic insight about the mechanisms leading to the disappearance of the nanostripes in favor of ferroelectric nanobubbles (that bear resemblance with ferromagnetic bubbles in ferromagnetic films [17]), and about the evolution of the morphology of these ferroelectric nanobubbles, under an electric field in BTO films – especially, when realizing that $BaTiO_3$ *bulk* is the most extensively studied system among all ferroelectrics [18] while little is still known about the domains [4] and their electric-field-induced domain evolution [9] in epitaxial $BaTiO_3$ *ultrathin films*.

The aims of this article are to use first-principles-based techniques to i) provide atomistic details of domains in BTO ultrathin films; ii) indicate how such domains evolve as a function of applied electric field; iii) compare the properties of BTO ultrathin films with those of PZT ultrathin films, as well as, to discuss and understand their similarities and differences.

The remainder of this paper is organized as follows. In Section II, the theoretical method is described. The stripe domains in 20Å-thick BTO thin films, and their energetic origins, are presented in Section III.A, with a special emphasis on explaining the difference in morphology between periodic nanostripes in BTO and PZT thin films. Section III.B depicts the electric-field-induced domain evolution in the BTO ultrathin films, and compares it with the corresponding domain evolution in PZT ultrathin films. Finally, a conclusion is provided in Section IV.

## II. THEORETICAL METHOD

The presently studied BTO and PZT ultrathin films are assumed to be grown along the [001] direction (z-axis) and are Ba-O and Pb-O terminated, respectively, at all surfaces. The PZT films are disordered in nature, with a Ti overall composition chosen to be 60%. Otherwise mentioned, the films with thickness of m lattice constants (which corresponds to a thickness of ~4m in Angstrom) are modeled using (relatively large) 24×24×m periodic supercells that are periodic *only* along the x- and y-axes (aligned along the pseudocubic [100] and [010] directions, respectively). The total energy of such supercells is used in Monte-Carlo (MC) simulations – that run over 80,000 sweeps – and is written as:

$$\varepsilon_{tot}(\{u_i\},\{v_i\},\eta) = \varepsilon_{Heff}(\{u_i\},\{v_i\},\eta) + \beta \Sigma_i <E_{dep}> \cdot Z^* u_i - \Sigma_i E \cdot Z^* u_i \quad (1)$$

where $u_i$ is the local soft mode in the unit cell $i$ of the film – whose product with the effective charge $Z^*$ yields the local electrical dipole in this cell; while $\eta$ and $\{v_i\}$ are the homogeneous strain tensor and inhomogeneous strain-related variables in unit cell $i$, respectively. [19] $\varepsilon_{Heff}$ represents the intrinsic (effective Hamiltonian) energy of ferroelectrics thin films. Its analytical expression is the one of Ref. 19 for BTO bulk and of Refs. 20 and 21 for PZT bulk (while its first-principles-derived parameters are those of Ref. 22 for BTO and of Refs. 20 and 21 for PZT), except for the dipole-dipole interactions for which the formula derived in Refs. 23 and 24 for *thin film under* ideal *open-circuit (OC)* conditions is used. [As a result, $\varepsilon_{Heff}$ of BTO consists of five energy parts − a local soft mode self energy ($\varepsilon^{self}$), an elastic strain energy ($\varepsilon^{elas}$), a short-range (up to third nearest neighbors) interaction ($\varepsilon^{short}$) between soft modes, a long range dipole-dipole interaction ($\varepsilon^{dpl}$), and an interaction between the local soft modes and local strain ($\varepsilon^{int}$) – while additional alloying terms are involved for the $\varepsilon_{Heff}$ of PZT]. Such electrical boundary conditions naturally lead to the existence of a *maximum* depolarizing field (denoted by $<E_{dep}>$ and determined from the atomistic approach of Ref. 23) inside the film, when the dipoles point along the [001] direction. The second term of Eq. (1) mimics a *screening* of $<E_{dep}>$ via the $\beta$ parameter. More precisely, the residual depolarizing field – resulting from the combination of the first and second term of Eq. (1) – has a magnitude equal to $(1-\beta)\,|<E_{dep}>|$. $\beta=0$ thus corresponds to ideal OC conditions, while an increase in $\beta$ lowers the magnitude of the resulting depolarizing field, and $\beta=1$ corresponds to ideal short-circuit (SC) conditions for which the depolarizing field has vanished. The third term of Eq. (1) represents the effect of a homogeneous electric field (***E***) on properties of the system. In the present study, we only consider an electric field applied along the *z*-axis (which corresponds to the realistic situation of thin films that are sandwiched by electrodes along [001]), and denote its magnitude as $E_z$. Note also that we mimic *epitaxially-grown* (001) films by freezing of some components of the $\eta$ homogeneous strain tensor, that are (in Voigt notation) $\eta_6=0$ and $\eta_1=\eta_2=\delta$ – with $\delta$ being the value forcing the film to adopt the in-plane lattice constant of the substrate. [6, 25, 26]

## III. RESULTS AND DISCUSSION

### A. Ferroelectric Nanodomains in Thin Films

We focus here on a 20 Å-thick (*m*=5) BTO film, as mimicked by a 24×24×5 supercell, under a compressive strain of −2.2% and having a realistic electrical boundary condition corresponding to $\beta$=0.8. [27] As consistent with Refs. 5, 6 and 9, the combination of a significant compressive epitaxial strain (that favors dipoles along the z-direction) with such electrical boundary condition (that will lead to a large enough residual depolarizing field if all the dipoles would parallely point along the z-axis) generates stripe domains at low temperature, when no external electric field is applied. Figure 1a displays such domains at 10K (that were obtained by slowly cooling down the system) and indicates that the stripe domains in the BTO thin films consist of periodically alternating "up" and "down" domains (here, the up (down) domains refer to the domains with the z-component of the local dipoles along the +z (–z) direction), as in compressively-strained (001) PZT and PbTiO$_3$ thin films. [5-7, 9-12] However, the stripe domains in the BTO film run along [110] and alternate along [1-10] – as also previously found in Ref. 4 – while the stripe domains in PZT and PbTiO$_3$ thin films run along [100] and alternate along [010] – as indicated in Refs. 5-7 and 9-11. Moreover, the periodicity of these "diagonal'' stripes in the 20 Å-thick BTO film is of ~4.3 lattice constants along [1-10] (which is consistent with Ref. 4) since Fig. 1a indicates an alternation of the stripe of 6 lattice constants along the [100] and [010] directions. On the other hand, we previously found in Refs. 6 and 9 that the periodicity of a PZT film with the same thickness (namely, 20 Å) is equal to 8 lattice constants along the [010] direction. To better understand why the stripes in the BTO film differ from those in the corresponding PZT film, Table I reports the supercell total energy, and its decompositions into the different terms of $\varepsilon_{Heff}$, at 10K, for three different stripe domains in 20 Å-thick BTO film (as all mimicked by a 24×24×5 supercell under identical boundary conditions): 1) the stripe domains depicted in Fig. 1a and that we will refer to as BTO-110-4.3; 2) the stripe domains alternating, as in PZT, along the

[010] direction with a period of 4 lattice constants (i.e., that is similar in length to the one of Fig. 1a) and that we will denote by BTO-010-4; and 3) the stripe domains also alternating along the [010] direction but now with a period of 8 lattice constants (i.e., as in the PZT film having similar thickness) and that we will refer to as BTO-010-8. Note that these two last domains were first created from scratch and then allowed to relax during a 80,000 sweeps of MC procedure at 10K. Interestingly, they were found to be thermodynamically (meta)stable at low temperature since the periodicity and directions of these stripe domains remained the same during the whole MC procedure at 10K.

Table I clearly reveals that the <110>-oriented stripes are energetically preferred over the <010>-alternating stripes in BTO films mainly because of long-range dipole-dipole interaction energy, $\varepsilon^{dpl}$, albeit at the cost of short-range energy, $\varepsilon^{short}$. Interestingly, the lower $\varepsilon^{dpl}$ in BTO-110-4.3 stripes is consistent with dipoles' configuration close to the domain walls: any of such dipoles, let's say located at site $i$, will interact with the two (respectively, one) antiparallel dipoles and with the two (respectively, three) parallel dipoles located at the four sites that are nearest neighbor (in the (001) plane) of site $i$, when the stripes alternate along [1-10] (respectively, [010]). This gain in number of nearest-neighbor *antiparallel* dipoles when going from stripes alternating along [010] to stripes alternating along [1-10] effectively lowers $\varepsilon^{dpl}$ − while raising $\varepsilon^{short}$ at the same time. In the case of BTO, Table I indicates that $\varepsilon^{dpl}$ is lowered more than $\varepsilon^{short}$ is raised, while we numerically found (not shown here) that the opposite occurs for PZT films because of the different parameters inherent to that latter material – which explains the difference in morphology of stripe domains in these two films.

As it can be seen in Figure 2, that shows the real-space distribution of the local dipoles in the ground-states, two other main differences exists between the morphology of the stripe domains in BTO versus PZT films. They are: (1) the dipoles of a given stripe domain are nearly constant in direction (i.e., parallel or antiparallel to the z-axis) and magnitude *inside* the BTO films, while such dipoles continuously rotate across the stripe *inside* the PZT films; (2) the dipoles at the *surfaces* can "only" deviate from the z-axis by up to 45 degrees in the BTO film, while the existence of *in-plane* surface dipoles were previously reported in PZT films in order to close the flux (and thus minimize depolarizing effects). [6, 9, 11] To better understand such differences, we decided to construct another domain pattern in a 20 Å-thick BTO film (keeping the same boundary conditions as above), to be denoted by $BTO_{PZT}$-010-8. Such latter state exhibits the same dipole configuration as the equilibrium domain pattern of a 20 Å-thick PZT film, but with the average magnitude of the dipoles having been rescaled (with respect to that of the PZT film) in order to be identical to that of BTO-010-8. As a result, $BTO_{PZT}$-010-8 exhibits continuously rotating dipoles across a given stripe and in-plane surface dipoles, as in the equilibrium domain pattern of PZT but unlike in any stable domain of BTO films discussed above (i.e., BTO-110-4.3, BTO-010-4 or BTO-010-8). Interestingly, unlike the BTO-010-4 and BTO-010-8 metastable stripe domains, $BTO_{PZT}$-010-8 is thermodynamically *unstable* because it directly transforms into BTO-010-8 after a couple of MC sweeps – which indicates that BTO films profoundly dislike significantly rotating and in-plane dipoles. Comparing the different energies of BTO-010-8 and $BTO_{PZT}$-010-8 stripe domains (see Table 1) reveals that such disliking takes a major part of its origin in the strain-soft mode interaction energy, $\varepsilon^{int}$. In other words, the local soft modes in BTO prefer to follow the easy polarization (z-) axis introduced by the elastic compressive strain. In PZT, we numerically found (not shown here) that $\varepsilon^{short}$ becomes the predominant contributing factor for the PZT domain patterns to exhibit rotating dipoles across stripes because dipoles can more easily rotate.

**B. Electric-Field-Induced Evolution of Ferroelectric Nanodomains**

We now turn our attention to the evolution of the BTO-110-4.3 ground-state domain when subject to an external electric field − for a chosen temperature of 10 K and with the boundary conditions being the same as those indicated above (i.e., a compressive strain of −2.2% and $\beta$=0.8). Consequently, Figure 3 shows the supercell average of the z-Cartesian component of the local modes ($<u_z>$) and of the *magnitude* of the local modes ($<u_M> = N^{-1} \Sigma_i [u_{i,x}^2 + u_{i,y}^2 + u_{i,z}^2]^{1/2}$, where N is the number of sites in the supercell) as a function of $E_z$ in the 20 Å-thick BTO ultrathin film. [For instance, Fig. 3 indicates that, when no electric field is applied ($E_z$=0), $<u_z>$ vanishes while $<u_M>$ exhibits a non-zero value, which is consistent with the BTO-110-4.3 domain pattern shown in Fig. 1a.] Atomistic detail on the evolution of the local dipoles under the applied electric field – as given by the last snap shot of our MC simulations – is given in Figs. 1a~1f. Moreover, Figures 4 display complementary statistical information on the number of supercell sites having positive or negative z-component of their dipoles, as well as, on the average magnitude of such dipoles in the different (001) planes, as a function of $E_z$.

As we will see, four regions can be distinguished, with their electric-field range being indicated in Figs. 3 and 4 for our film of interest (Note that such range depends on the films thickness and boundary conditions [9]).

**Region I:** Figs. 4 tell us that, when the electric field is applied from zero to $33\times10^7$ V/m (Region I), the local dipoles in the "down" domains do *not* switch their direction to point along the +z direction (i.e., parallel to the electrical field) but rather "simply" significantly decrease their magnitude – while the dipoles in the "up" domains more modestly increase in magnitude. Such behaviors lead to $<u_z>$ and $<u_M>$ increasing and slightly decreasing with $E_z$, respectively, (see Fig. 3) and to domain walls not moving with respect to the $E_z$=0 case (as shown in Fig. 1b). Interestingly, Region I (unlike the Regions II, III and IV that will be discussed below) was not found in PZT films subject to an electric field, [9] because of the easiness in rotating dipoles in these latter systems.

**Region II:** A further increase of $E_z$ from $33\times10^7$ to $43\times10^7$ V/m (Region II) induces a switch, from the −z to +z direction, of some dipoles in the down domains − as indicated by Figs. 4a and 4c. Such switch thus results in an increase of $<u_z>$ when increasing $E_z$ (see Fig. 3) and in an expansion of the majority domains at the expense of minority domains while maintaining the same overall stripe periodicity – as in the corresponding region for PZT films. [9] However, unlike in PZT [9] and as seen in Fig. 1c, this expansion of majority domains is *not* synchronous throughout the domain walls. Rather, scattered regions of switched dipoles (with five lattice constants along the z-axis and one lattice constant along both the *x*- and *y*-axes) that are bounded by (100) and (010) side surfaces "pop out" adjacent to the original domain walls. Therefore, the domain walls expand along [100] and [010] by one lattice constant at a time, and now exhibit a zigzag pattern (while these domain walls are straight in PZT films [9]).

Moreover, the switching behaviors of the dipoles on surfaces and inside the films are documented in Figs. 4. In particular, one can see that the dipoles on surfaces are slightly easier to switch than the dipoles inside the films because they have smaller magnitude just before the switching. The real-space distribution of dipoles (data not shown) in Region II further indicates that the switching of the *surface* dipoles occurs through both rotation and flipping mechanisms. The former mainly involves the dipoles that initially deviate from the *z*-axis (see Fig. 2) and is accomplished by local dipoles rotating away from the z-axis while decreasing their magnitude at the same time. The latter applies to the dipoles that were initially aligned along the −z-direction (see Fig. 2) is preceded by a decreasing of the dipoles' magnitude and leads to a significant change of this magnitude after the flipping (see Figs. 4b and 4d). In contrast, the dipoles *inside* the films switch predominantly through flipping (see Figs. 1a-1c), with the direction being reversed and the magnitude of dipoles being only very slightly increased after such flipping (see Figs. 4b and 4d). In other words, the slight increase of <$u_M$> when increasing $E_z$ in Region II (see Fig. 3) mostly originates from the surface dipoles.

**Region III:** As $E_z$ keeps increasing, the minority (down) stripe domains become pinched along [110]. Nanobubbles now form (see Fig. 1d), as in PZT thin films under electric field. Note that the stripe to bubble domain transition is numerically found for $E_z$ in between $41\times10^7$ and $45\times10^7$ V/m (because of the zigzagged domain walls, it is difficult to exactly pin point the onset of stripe to bubble domain transition, unlike in PZT thin films [9]). Once the nanobubbles are formed, they dramatically contract along [110] while keeping their width along [1-10] more or less the same under further applied field − as seen in Figs. 1d and 1e. In other words, as in PZT films, [9] the bubbles contract parallely to the direction along which the initial stripe domains were running (i.e., [110] in BTO *versus* [100] in PZT films) while keeping their width fixed parallel to the direction along which the initial stripe domains were alternating (i.e., [1-10] in BTO *versus* [010] in PZT ultrathin films). Interestingly, Fig. 4d reveals that, between $43\times10^7$ and $51\times10^7$ V/m, the average magnitude of the *surface* dipoles having a positive component along the direction of the field (i.e., with $u_z>0$) *decreases* with $E_z$. [This is due to the facts that the number of switched surface dipoles (i.e., having now $u_z>0$ while initially exhibiting $u_z<0$) increases rather rapidly with $E_z$ (see Fig. 4c) while such dipoles had a rather small magnitude before this switching (see Fig. 4b).] Figs. 4a further indicates that the critical field for which *all* the surface dipoles belonging to the initial down stripes have *completely* switched is around $E_z=61\times10^7$ V/m (Note that *half* of the dipoles inside the film that belonged to the initial down domains are still not flipped under such field, according to Fig. 4a). In other words, the part of Region III located above $E_z=61\times10^7$ V/m possesses nanobubbles that are *encapsulated* (i.e., that do not touch the surfaces, as in PZT films). With further increase of $E_z$ (see Fig. 1e), the bubbles still continue to contract along [110]. Once they reach a critical size of two lattice constants along both *x*- and *y*- axes and of three lattice constants along the *z*-axis, some bubbles pop out and thus the number of nanobubbles decreases with $E_z$. Note that we numerically found that the collapse of BTO bubbles is via flipping dipoles inside the films, while the collapse of PZT bubbles is via dipole rotation.

**Region IV:** At $E_z=104\times10^7$ V/m, all the nanobubbles have disappeared and a single monodomain, with all the dipoles pointing along the +z direction, is thus formed − as seen in Fig. 1f and as consistent with the fact that Fig. 3 indicates that $<u_z>$ and $<u_M>$ are now identical. In this monodomain state, all the dipoles increase in magnitude with $E_z$, with the surface dipoles being larger in size.

## IV. CONCLUSION

In summary, using a first-principles-derived method, we studied the morphology of periodic stripe nanodomains, and their evolution under an applied electric field, in epitaxial (001) BTO ultrathin films. It is predicted that, under zero field, such stripes alternate along [1-10] (as consistent with Ref. 4) while they are known to alternate along [010] in compressively-strained PZT and $PbTiO_3$ ultrathin films. [5-7, 9-11] This difference is traced back to the different balance between dipole-dipole interactions and short-range energies in BTO versus PZT thin films. Moreover, no in-plane dipoles exist at the surface of BTO films, unlike what was predicted in PZT two-dimensional objects [6, 9] − mostly because of energy terms related to the coupling between strain and dipoles. (Dipoles can deviate from the [001] direction by "only" up to 45 degrees at the surfaces of the BTO films). Furthermore, the stripe→bubble→monodomain transition sequence recently predicted in PZT ultrathin films [9] also occurs in epitaxial BTO ultrathin films, but with some noteworthy differences. For instance, for BTO films, (1) the dipoles in the minority domains films first decrease in magnitude before switching their directions; (2) zigzagged domains walls exist in the stripe regime; (3) bubbles contract along [110], rather than [100], before "popping out". We hope that our predictions will be confirmed soon, and lead to a deeper knowledge of ferroelectric thin films.

**ACKNOWLEDGMENTS**

This work is supported by DOE Grant No. DE-FG02-05ER46188, ONR Grants Nos. N00014-04-1-0413, N00014-01-1-0365 (CPD), and N00014-01-1-0600, and by NSF Grant Nos. DMR-0404335 and DMR-0080054 (C-SPIN). L.B. thanks the twenty-first century professorship in nanotechnology and science education

Table I: Energies, in atomic units, of the 24×24×5 supercell mimicking the different BTO stripe domains (see text). The positive sign of the total energy originates from the epitaxial compressive strain.

| Energy (Ha) | BTO-110-4.3 | BTO-010-4 | BTO-010-8 | BTO$_{PZT}$-010-8 |
|---|---|---|---|---|
| $\varepsilon^{self}$ | 13.076 | 12.738 | 10.574 | 11.082 |
| $\varepsilon^{elas}$ | 5.401 | 5.347 | 4.795 | 4.795 |
| $\varepsilon^{int}$ | -6.126 | -5.996 | -3.710 | -3.065 |
| $\varepsilon^{short}$ | 5.763 | 4.774 | 3.516 | 3.043 |
| $\varepsilon^{dpl}$ | -16.279 | -14.930 | -12.777 | -12.420 |
| $\varepsilon_{tot}$ | 1.836 | 1.932 | 2.399 | 3.436 |

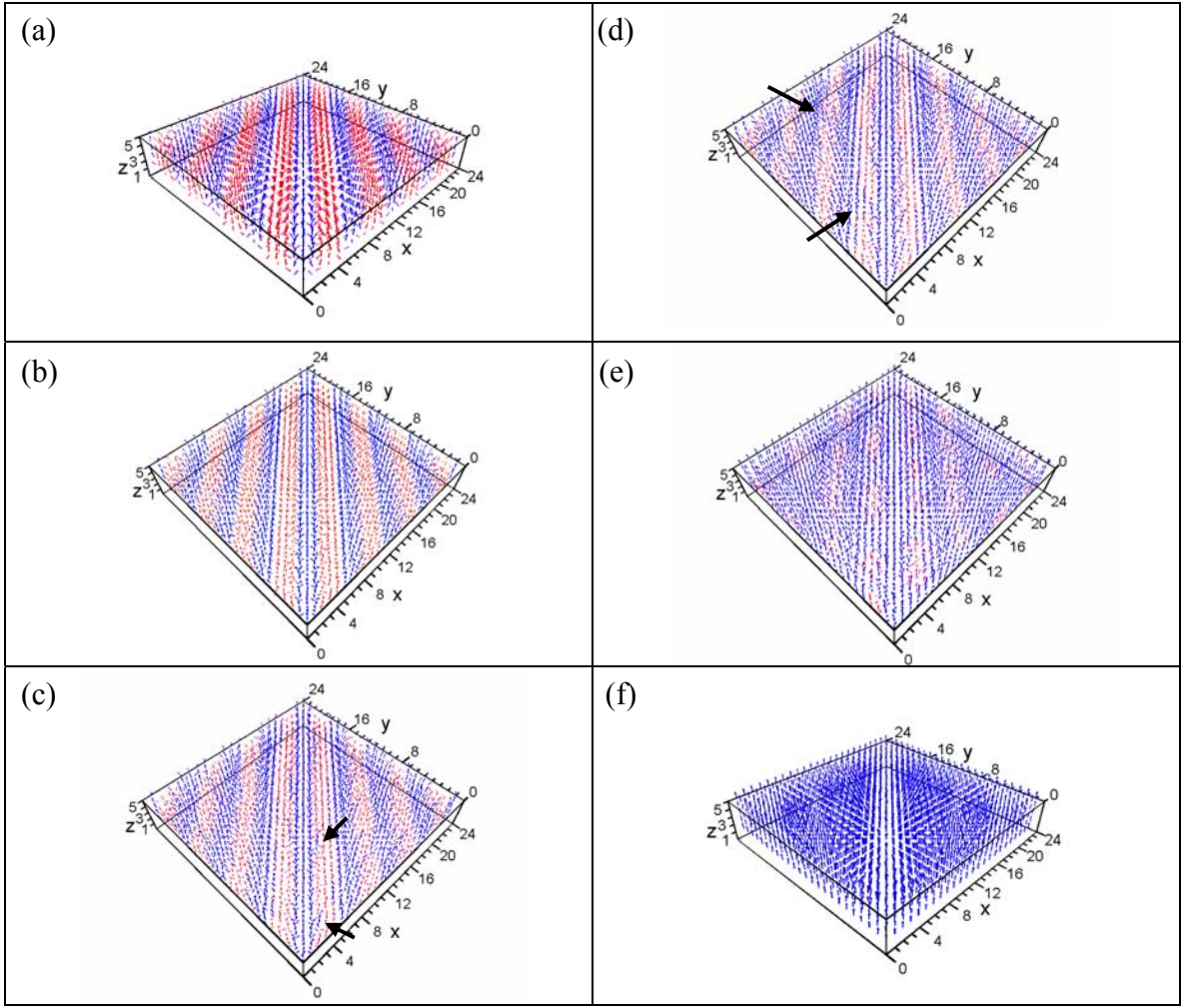

Fig.1. (color online) T=10K three-dimensional polarization patterns in 20 Å-thick BTO (001) films under a compressive strain of -2.2% and a 80% screening of the maximum depolarizing field, for different $E_z$: stripe domains under (a) $E_z=0$, (b) $E_z=30\times10^7$ and (c) $E_z=41\times10^7$ V/m; bubble domains under (d) $E_z=51\times10^7$ and (e) $E_z=71\times10^7$ V/m; and monodomain under (f) $E_z=104\times10^7$ V/m. Blue (red) arrow characterizes local dipoles having a positive (negative) component along the *z*-axis. The arrows in (c) and (d) indicate some representative switched dipoles (that lead to zigzagged domain walls) and pinching of stripes (that generates nanobubbles), respectively.

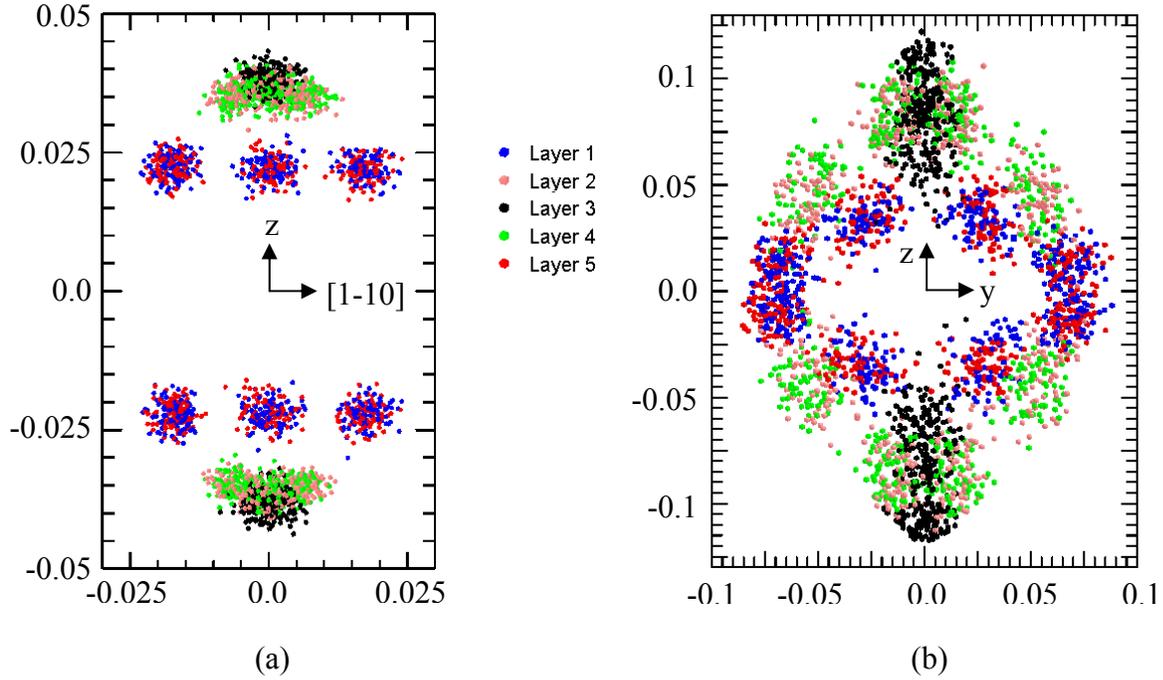

Fig.2. (color online) Real-space distribution of the local modes in (a) BTO-110-4.3 and (b) PZT-010-8 films. These films possess 5 (001) B-layers (i.e. m=5), or equivalently, have a thickness of around 20 Å. The blue and red symbols correspond to the surface B-layers, while the black symbol characterizes the dipoles at the most inner layer. The green and orange symbols refer to the other two (001) planes (located in between the surface and most-inner layers). Vertical and horizontal axes (in both (a) and (b) parts) represent the projection of the local modes (given in atomic units) along the $z$-axis and stripe's alternating direction, respectively.

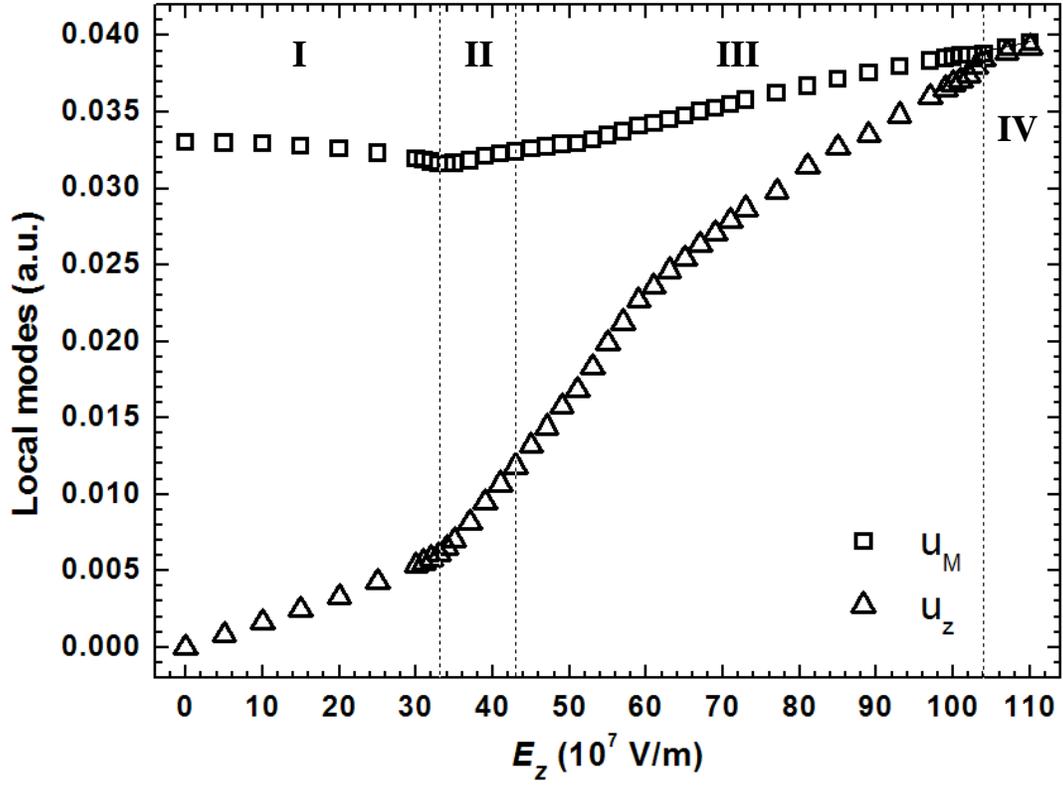

Fig.3. Supercell average of the $z$-Cartesian component of the local modes, $\langle u_z \rangle$, and supercell average of the magnitude of the local modes, $\langle u_M \rangle$, as a function of $E_z$ for 20 Å-thick BTO films under a compressive strain of -2.2%, a 80% screening of the maximum depolarizing field, and at $T=10$ K. Square and triangle symbols display $\langle u_M \rangle$ and $\langle u_z \rangle$, respectively. The supercell averages of the $x$- and $y$-Cartesian components of the local modes are essentially null and are not shown for clarity.

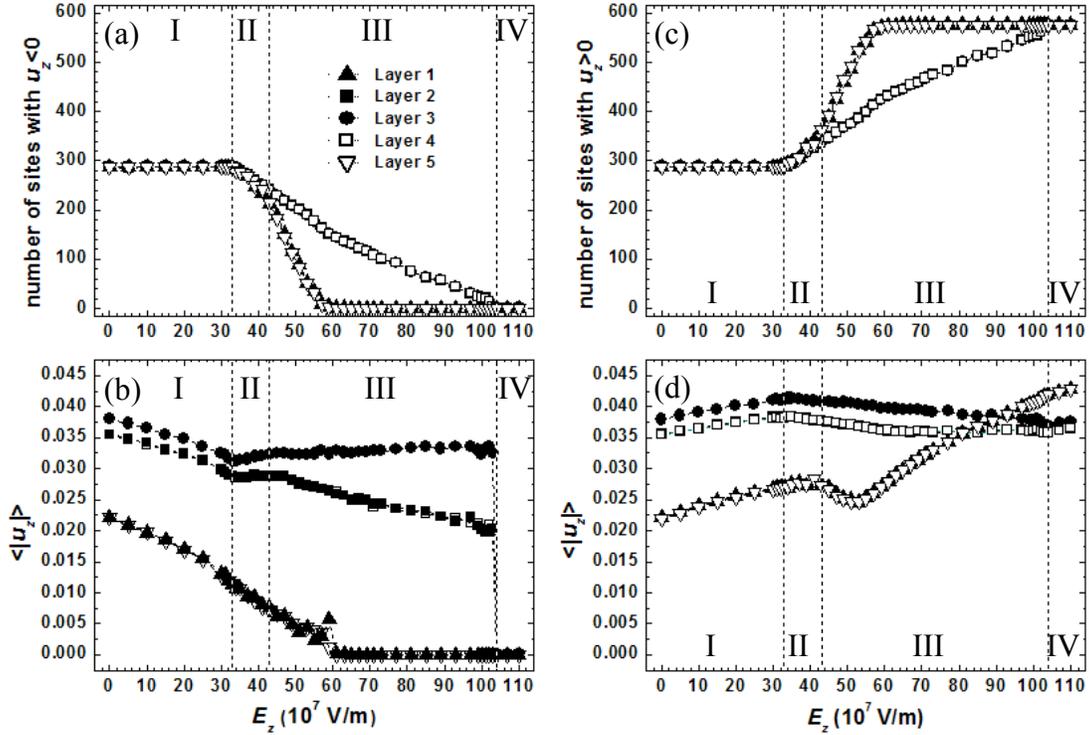

Fig.4. (a) Number of sites with a negative $z$-Cartesian component of the local modes and (b) the corresponding $\langle|u_z|\rangle$ average magnitude of their $z$-Cartesian component of the local modes, as a function of $E_z$, in the different (001) layers at $T=10$ K for 20 Å-thick BTO films under a compressive strain of -2.2 and a 80% screening of the maximum depolarizing field. (c) and (d) are the same as (a) and (b), respectively, but for a positive $z$-Cartesian component of the local modes. The layer index of thin films from the first layer (layer 1, which is a surface layer) to the last layer (layer 5, which is the other surface layer) are indicated via solid triangle, solid square, solid diamond, open square and reversed open triangle, respectively.